\documentclass[sigconf, nonacm]{acmart}

\newcommand\vldbdoi{XX.XX/XXX.XX}
\newcommand\vldbpages{XXX-XXX}
\newcommand\vldbvolume{17}
\newcommand\vldbissue{1}
\newcommand\vldbyear{2023}
\newcommand\vldbauthors{\authors}
\newcommand\vldbtitle{\shorttitle} 
\newcommand\vldbavailabilityurl{https://github.com/gtfintechlab/Universal-NFT-Vector-Database}
\newcommand\vldbpagestyle{empty} 

\begin{document}
\title{The Universal NFT Vector Database: A Scaleable Vector Database
for NFT Similarity Matching}

\author{Samrat Sahoo}
\affiliation{%
  \institution{Georgia Institute of Technology}
  \city{Atlanta}
  \state{Georgia}
}
\email{ssahoo61@gatech.edu}

\author{Nitin Paul}
\affiliation{%
  \institution{Georgia Institute of Technology}
  \city{Atlanta}
  \state{Georgia}
}
\email{npaul34@gatech.edu}

\author{Agam Shah}
\affiliation{%
  \institution{Georgia Institute of Technology}
  \city{Atlanta}
  \state{Georgia}
}
\email{ashah482@gatech.edu}

\author{Andrew Hornback}
\affiliation{%
  \institution{Georgia Institute of Technology}
  \city{Atlanta}
  \state{Georgia}
}
\email{ahornback6@gatech.edu}

\author{Sudheer Chava}
\affiliation{%
  \institution{Georgia Institute of Technology}
  \city{Atlanta}
  \state{Georgia}
}
\email{sudheer.chava@scheller.gatech.edu}

\begin{abstract}
Non-Fungible Tokens (NFTs) are a type of digital asset that represents a proof of ownership over a particular digital item such as art, music, or real estate. Due to the non-fungible nature of NFTs, duplicate tokens should not possess the same value. However, with the surge of new blockchains and a massive influx of NFTs being created, a wealth of NFT data is being generated without a method of tracking similarity. This enables people to create almost identical NFTs by changing one pixel or one byte of data. Despite the similarity among NFTs, each NFT is assigned a completely different token ID. To address the NFT duplication issue, we developed a modular, easily-extendable, hardware-agnostic, cloud-centered NFT processing system that represents NFTs as vectors. We established a database containing a vector representation of the NFTs in accordance with the Ethereum Request for Comment 721 (ERC-721) token standards to initiate the process of aggregating NFT data from various blockchains. Finally, we developed an NFT visualization dashboard application with a user-friendly graphical user interface (GUI) to provide non-technical users access to the aggregated NFT data. The Universal NFT Vector Database is an off-chain framework for NFT data aggregation based on similarity, which provides an organized way to query and analyze NFT data that was previously unavailable through on-chain solutions.
\end{abstract}

\maketitle

\pagestyle{\vldbpagestyle}
\begingroup\small\noindent\raggedright\textbf{PVLDB Reference Format:}\\
\vldbauthors. \vldbtitle. PVLDB, \vldbvolume(\vldbissue): \vldbpages, \vldbyear.\\
\href{https://doi.org/\vldbdoi}{doi:\vldbdoi}
\endgroup
\begingroup
\renewcommand\thefootnote{}\footnote{\noindent
This work is licensed under the Creative Commons BY-NC-ND 4.0 International License. Visit \url{https://creativecommons.org/licenses/by-nc-nd/4.0/} to view a copy of this license. For any use beyond those covered by this license, obtain permission by emailing \href{mailto:info@vldb.org}{info@vldb.org}. Copyright is held by the owner/author(s). Publication rights licensed to the VLDB Endowment. \\
\raggedright Proceedings of the VLDB Endowment, Vol. \vldbvolume, No. \vldbissue\ %
ISSN 2150-8097. \\
\href{https://doi.org/\vldbdoi}{doi:\vldbdoi} \\
}\addtocounter{footnote}{-1}\endgroup

\ifdefempty{\vldbavailabilityurl}{}{
\vspace{.3cm}
\begingroup\small\noindent\raggedright\textbf{PVLDB Artifact Availability:}\\
The source code, data, and/or other artifacts have been made available at \url{\vldbavailabilityurl}.
\endgroup
}
\begin{figure*}
  \includegraphics[width=0.60\textwidth]{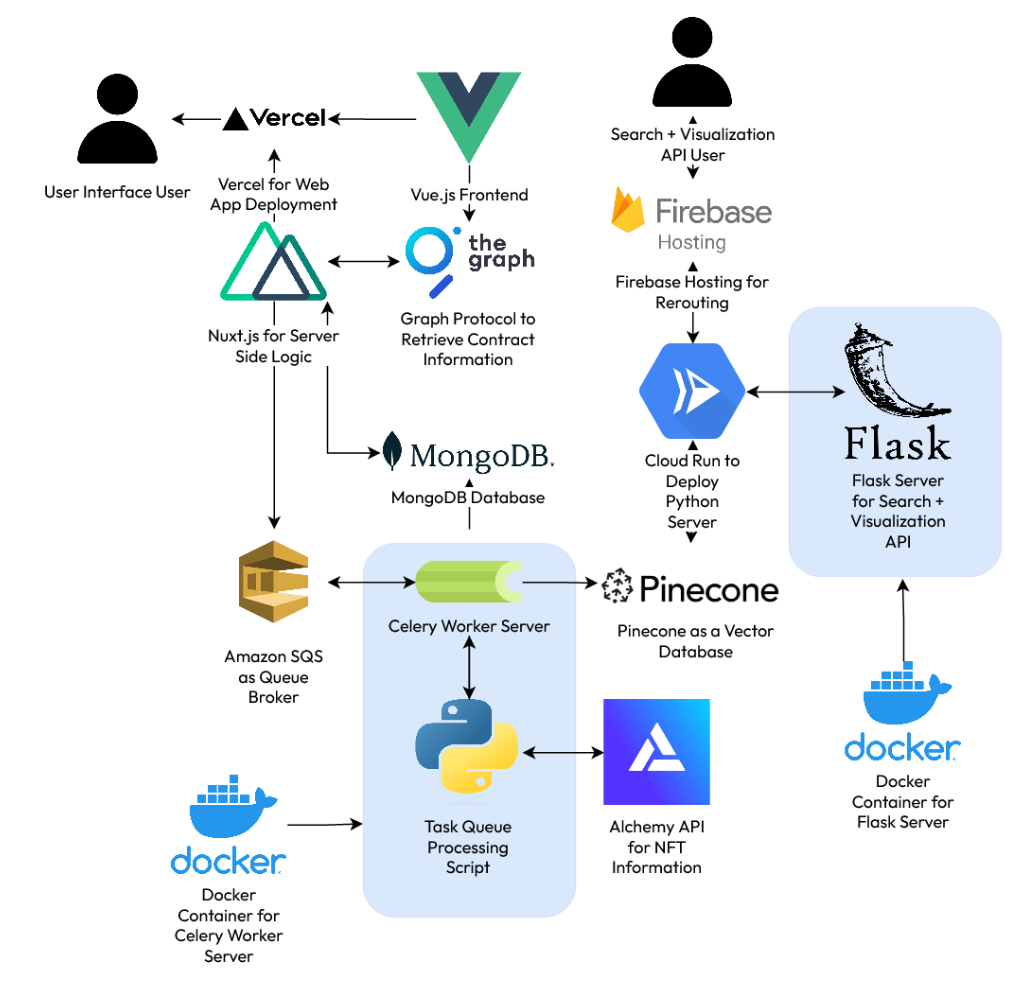}
  \caption{\normalfont This figure illustrates the infrastructure of the system we have designed. The upper left illustrates our client and server-side logic and blockchain data collection tools. The bottom represents the task queue processing system which processes NFT collections and NFTs. The upper right shows the infrastructure for our search and visualization APIs.}
  \label{fig:system_design}
\end{figure*}

\section{Introduction}

NFTs are digital assets on the blockchain that are intertwined with cryptography, containing metadata and a distinct hash for identification \cite{Wang21}. Designed to be cryptographically unique, NFTs originated from the Bitcoin blockchain in 2012 and have become widespread on the Ethereum blockchain \cite{Assia15}. Typically represented with the ERC-721 standard \cite{Entriken18}, NFTs are ideal for use cases involving collectible items, digital key management, concert tickets, and more because they establish uniqueness among digital tokens. To enforce authenticity between NFTs and improve fairness in the NFT ecosystem, especially among image-based NFTs, researchers and blockchain developers are searching for mechanisms to detect and enforce the cryptographic uniqueness enabling blockchain systems and infrastructure to fulfill their purpose \cite{Birch22}. 

Currently, it is expensive to store raw file data on a blockchain. As a result, NFTs on a blockchain store a reference (i.e., a URL) to the file data, and off-chain storage solutions host the metadata. The NFT is assigned a unique token ID; while the token ID is unique, the off-chain data may not be. With recent blockchain infrastructure development, there is a lack of robust protocols and standards to enforce the authenticity of an NFT. This means that an entity can duplicate the content of the original NFT, change one byte, and create a completely new NFT. As a result, the lack of enforcement technology impedes new digital innovations which rely on NFTs to be unique and not be replicated by another entity, effectively nullifying the original intention of NFT technology.

To address the challenges with NFTs, we propose a modular, easily-extendable, hardware-agnostic, cloud-centered NFT database. Such infrastructure helps enable and facilitate the enforcement of token uniqueness. It standardizes the data to a vector data repository which users can easily access through a web interface for research and utility purposes. The vector database determines similarities between NFT feature vectors to conclude the uniqueness of an NFT. 

First, to develop the NFT vector database, we extract the ERC-721 contracts using the Graph Protocol. After receiving the contract information, we collect the individual NFT data through Alchemy (a blockchain infrastructure provider). From the media URL associated with the NFT data, we aggregate the images, embed them, and place the vectors (each containing a distinct ID) in Pinecone, a vector database provider. 

Next, we connect the database to an interactive GUI to derive updated metrics and critical information regarding NFT contract data. The web application enables researchers and consumers to interact with the vector database and have a direct access point to the NFT data. Furthermore, this infrastructure for NFT data could be generalized for various applications surrounding the necessity of unique digital assets, such as identity authentication and NFT filtration systems \cite{Mecozzi22}.

Through this study, we present a framework for enforcing the uniqueness of NFTs to extract more value from these tokens to be used by researchers and consumers. Additionally, We produce a web application interface with the NFT vector database with a user-friendly graphical user interface to improve access to NFT data and advance future innovations involving NFTs. This study aims to illustrate the system's extensibility and modular design, enabling developers to replace parts of the system instead of developing new infrastructure for each application area.

\section{Data Collection and Processing}
\subsection{Blockchain Data Querying and Collection}
Ethereum blockchain data is challenging to query due to the blockchain’s inherent linked-list structure. Traditionally, querying the blockchain for all ERC-721 contracts and tokens would require spinning up an Ethereum RPC node and iterating through every block. Our system, instead, leverages the Graph Protocol, a decentralized querying layer built on top of Ethereum \cite{Yaniv19}, allowing for easy and fast querying of subsets of blockchain data (called subgraphs) via a GraphQL API. We utilize an EIP-721 subgraph \cite{Sandford19} that monitors all ERC-721 contracts and collects contract addresses through this subgraph. We collect these contract addresses in a paginated manner -- processing anywhere from 10 to 100 contracts at a time -- to ensure neither the subgraph infrastructure nor our infrastructure is overwhelmed with excessive data input and output operations. 

We tested two methods to obtain the individual NFT data for each ERC-721 contract. The first method utilized the subgraph used for the contracts to acquire the NFTs and individual details of each NFT. The second method involved integrating the system with Alchemy’s -- a Web3 infrastructure provider -- NFT infrastructure. Since speed was a priority for us in retrieving as many NFTs as possible, we ran benchmark tests to measure which method would run faster. We discovered that Alchemy’s NFT infrastructure had significant time improvements and scales roughly linearly according to the number of NFTs. The linear scaling is because Alchemy caches NFT media information in a database, allowing almost instant retrieval. 

\subsection{Data Processing}
By utilizing Alchemy’s NFT Infrastructure, we obtain the media URL of an individual NFT. We feed this media URL through a data processing system that obtains the file contents at the URL. Our system conducts no preprocessing steps on the media contents to retain the inherent attributes of the NFT. The image is then directly converted into a 2016x1 dimensional vector via a RegNetY-080 embedding \cite{Radosavovic20}. The system divides the data into two database pipelines. The first pipeline stores the generated vector in a vector database provider called Pinecone. The second pipeline stores additional information about the NFT, such as chain information, the NFT metadata URL, and NFT standard information in a MongoDB database. The dual database system enables users to obtain NFT details (from the MongoDB database) based on visual aspects of NFTs (using the vector database). Our design differs from traditional methods, which rely solely on querying based on the NFT collection contract ID and token ID.

\subsection{Vector Operations and Dimensionality Reduction}
Raw images have large, arbitrary dimensions when converted to a vector, posing two issues our system attempts to solve. First, large dimensions are storage intensive, with even smaller images requiring tens of thousands of elements per vector. By utilizing image embeddings, we can apply dimensionality reduction to the image while retaining the inherent attributes of the NFT. The RegNetY-080 represents the image as a 2016x1 dimensional vector, significantly reducing the amount of space each image takes up in the database. 

The second problem of arbitrary vector dimensions poses an issue for vector comparison. Current metrics for vector comparison, such as euclidean distance, require consistently sized vectors \cite{Dokmanic15}. However, with different-sized images, it is impossible to compare the images without some standardization. After standardizing the vectors, we were able to add functionality to search for similar NFTs -- we chose to use cosine distance as our search metric as cosine distance tends to perform best for higher dimensional embeddings \cite{Schubert21}. 

For the visualization capabilities, we need an effective dimensionality reduction technique to reduce the embedding vectors into a two-dimensional surface. To determine which dimensionality reduction technique is most effective, we run experiments to compare different methods including multidimensional scaling, t-distributed stochastic neighbor embedding, principal component analysis, truncated singular value decomposition, and isometric mapping. For this experiment, we define and calculate a metric called the cluster ratio to determine which low-dimensionality reduction technique is most effective -- a higher cluster ratio indicates a more effective dimensionality reduction technique. We define effectiveness as the algorithm that maximizes the distance between NFT collections clusters centers while minimizing the distance between the individual NFTs and the NFT collection cluster centers. We define cluster ratio, cluster distance, and collection distance as follows:
\begin{itemize}
\item \bf Cluster Distance: \normalfont This is the Euclidean distance between the centers of two clusters of NFTs.
\item \bf Collection Distance: \normalfont This is the average Euclidean distance between the individual elements of each NFT cluster and the center of the NFT cluster.
\item \bf Cluster Ratio: \normalfont This is the ratio of the distance between two cluster centers and the average distance between the individual elements of each cluster and the respective cluster center of the elements. The cluster ratio is better defined as the ratio of cluster distance to collection distance.
\end{itemize}

We used an NFT dataset with 50 NFTs from two different NFT collections. Each NFT collection has NFTs with similar aesthetic attributes. The initial step in determining the best algorithm was creating embeddings for all of the images in the dataset. We then applied the dimensionality reduction technique we were testing and created two clusters on the lower dimensional data using a K-means clustering algorithm. From this, we calculated the cluster distance, collection distance, and cluster ratio. The truncated singular value decomposition had the higher cluster ratio and was the chosen dimensionality reduction technique for the visualization capabilities.

\section{Software Infrastructure}
\subsection{Control Panel, Dashboard, and Backend Server}
Our GUI grants nontechnical users complete access to the system’s capabilities. To create this GUI, we leveraged a standard software engineering model for building applications called the Model-View-Controller (MVC) pattern to ensure organized, secure, and easy access to data:
\begin{itemize}
\item \bf Model:  \normalfont The model portion of the application primarily entails the database schemas, which enable a better organization of data -- we write and enforce using the Mongoose client for MongoDB when inserting data into our database. Additionally, using a strongly-typed language like TypeScript allows us to declare and enforce types strictly. 
\item \bf View: \normalfont The view portion of the application enables users to view current data and perform create, read, update, and delete (CRUD) operations, all abstractedly. In the case of the GUI, users leverage CRUD operations through actions such as viewing analytics, controlling the task queue, and updating analytics.
\item \bf Controller: \normalfont The system's controller is a backend server that manages all CRUD operations' logic. The CRUD operations include actions like retrieving analytics and updating the task queue. Additionally, the controller enables greater security in the application's admin panel, where users are authenticated via JavaScript Object Notation (JSON) web tokens, allowing them to control the task queue. The controller was written in TypeScript using a full-stack web framework known as Nuxt.js, enabling client-side and server-side under a mono repository structure, greatly simplifying the organization and infrastructure of the project. 
\end{itemize}

\subsection{Task Queue, Workers, and Horizontal Scalability}
Converting an image into a vector using image embeddings is computationally intensive. As a result, loading this onto an API would cause an overload of requests, requiring vertical scalability (increasing computational resources) and resulting in a less robust image processing system. To solve this, we implement a task queue system in which NFT data can be saved and processed on a separate server, allowing horizontal scalability. The flow of this system starts at the admin dashboard, where authenticated users can add NFT collections to the task queue. When the admin adds these collections, the system stores supplementary information regarding the processed item along with an ID and status in a MongoDB task queue collection. The system sends the ID to a messaging queue. We use Amazon Web Services (AWS) Simple Queue Service (SQS) as the messaging queue. These items are then individually processed on a separate server running a Celery worker service with two different tasks:

\begin{itemize}
\item \bf Contract Task:  \normalfont The first task is a contract task where the workers will get all the individual NFTs for a certain contract using Alchemy’s NFT infrastructure and add task items into the task queue for each NFT as discussed in section 2.2. 
\item \bf NFT Task: \normalfont The second task is an NFT task where the workers will get the data at the media URLs and convert this to a 2016 dimensional vector via the RegNetY-080 embedding as discussed in section 2.2. 
\end{itemize}
This architecture allows the system to be agnostic to a machine's physical hardware and focuses on reusability across devices.

\subsection{Search and Visualization API}
We offer a public search and visualization API to enable users to interact with the vector database. Both APIs also have interactive user interfaces in our web application.
\begin{itemize}
\item \bf Search API: \normalfont The search API will take any base64 encoded image and find the closest NFTs to the image by taking the cosine distance between the source image vector -- derived from the RegNetY-080 image embedding -- and the vectors in the database. We utilize Pinecone’s vector database search API to search for the appropriate vectors.  
\item \bf Visualization API: \normalfont The goal of the visualization API is to apply dimensionality reduction on a set of vectors via a t-distributed stochastic neighbor embedding. The visualization API will take in a list of vectors and reduce them to two dimensions. Each vector in the newly produced vectors corresponds to a point on a cartesian coordinate system.
\end{itemize}
\subsection{System Deployment}
Deploying with a combination of cloud and on-premise servers enables us to scale and process more NFTs per second on dedicated, robust hardware. When deploying the system, the objective was to keep costs low and the processing efficiency high. We had to deploy five systems: the vector database, MongoDB database, task queue, public APIs, and web application. Figure \ref{fig:system_design} details the deployment infrastructure.

\begin{itemize}
\item \bf Vector Database: \normalfont By using the Pinecone vector database, we greatly simplify the architecture by avoiding manually configuring scaling, storage, sharding, and other infrastructure details.
\item \bf Task Queue: \normalfont For the task queue processing system, we deploy two different systems. The first system was the task queue, which stored the other task IDs. For this, we utilized Amazon Simple Queue Service, which would act as the broker for our Celery worker server. The second is the Celery worker server which is set up as a system service in a Docker container and deployed on an on-premise server where it runs 24/7 and is always waiting to process task queue items. 
\item \bf Public APIs: \normalfont For the public APIs, we dockerize and deploy them to Google Cloud Run as serverless functions. We use Firebase hosting to direct HTTP requests to trigger functions in our containerized application. Following a serverless architecture, the container horizontally scales up or down depending on API traffic while ensuring that the costs match the usage.
\end{itemize}

\section{Conclusion}
In this study, we designed and developed a framework for enforcing the uniqueness of NFTs using a system with an NFT vector database as its central component to improve the value that users can extract inherently from NFTs. We developed an NFT vector database by extracting contract information and then querying the individual NFT data to find the media URL from which we aggregated the NFT images and placed them accordingly as vectors in the database. We also improved access to NFT data and associated metrics by allowing easy access through an intuitive web application interface where users can search for NFTs in the database and receive similarity scores. Our data access and standardized framework solution for NFT uniqueness based on similarity help make the NFT ecosystem fairer for blockchain developers, consumers, and researchers to leverage its potential for future innovations and research across various applications and fields.


\bibliographystyle{ACM-Reference-Format}
\bibliography{sample}

\end{document}